# Preparation of 18-filament Cu/NbZr/MgB$_2$ tape with high transport critical current density

C.F.Liu, S.J.Du, G.Yan, Y.Feng, X.Wu, J.R.Wang, X.H.Liu, P.X.Zhang, X.Z.Wu, L.Zhou

Northwest Institute For Nonferrous Metal Research, P.O.Box 51, Xi'an, PRC

L.Z.Cao, K.Q.Ruan, C.Y.Wang, X.G.Li, G.E.Zhou, Y.H.Zhang

University of Science & Technology of China, Hefei, PRC

**ABSTRACT** Cu-stabilized 18-filament MgB$_2$ tapes with NbZr buffer have been fabricated by using powder-in-tube technique. The phase composition and microstructure were examined by X-ray diffraction and optical microscopy. Transport critical current measurements were performed by a standard four-probe technique at different magnetic fields and temperature. The sample shows a high transport critical current density of 80,000 A/cm$^2$ (10K, 0 T) and 13,600 A/cm$^2$ (10K, 1T).

## 1. Introduction

The recent discovery of superconductivity at 39 K in the MgB$_2$ compound by Nagamatsu et al.[1] has triggered a great interest of the researchers in applied superconductivity[2]. Several groups have reported the transport critical current densities about $10^4$-$10^5$ A/cm$^2$ at 4.2K in single-filament MgB$_2$ tapes[3-7]. Recently, Soltanian et al.[8] have shown that promising transport critical current densities up to 16,000 A/cm$^2$ at 29.5K in 1T and 33K in null field can be achieved in Fe sheathed single-filament MgB$_2$ tapes. Although powder-in-tube (PIT) method is promising to obtain high quality MgB$_2$ wires and tapes, it is necessary to optimize this process in order to further improve superconducting properties. Also, multifilament wires and tapes are of importance for engineering application. In this work we first report high transport critical current densities in 18-filament MgB$_2$ tapes.

## 2. Experiment

Standard PIT methods were used for fabrication of the multifilament MgB$_2$ tape. Mg and B powder with a nominal stoichiometry of MgB$_2$ were well mixed and filled into a NbZr alloy tube with an outside diameter of 6mm and a wall thickness 1.5mm in order to prevent the reaction between Mg and Cu. Then the filled tube was inserted into a copper tube, having an inside diameter of 6.2mm with a wall thickness 2mm. The composite tube was swaged and drawn down to a 3×3 mm square wire with an intermediate annealing. The square wire was subsequently rolled into tape over many steps. Several short samples with 3 cm in length were cut from the tape and sintered in a tube furnace over a temperature range of 600-1000℃ for 1-10h in a high purity argon gas throughout the sintering process. The phase composition and microstructure were examined by X-ray diffraction and optical microscopy. Transport critical current has been measured by the standard four-probe technique.

## 3. Results and discussion

The X-ray diffraction result of the Cu/NbZr/MgB$_2$ tape is shown in Fig.1. It can be observed that the main phase is MgB$_2$ with a very small amount of MgO. Fig.2 illustrates the optical photography for a cross section of the sample. By using NbZr as a buffer layer, almost no reaction between MgB$_2$ and sheath was found. The total area of superconducting cross section of tape calculated from the optical photography is $0.0366mm^2$. The V-I curves of $Cu/NbZr/MgB_2$ tape at 10-35K in zero field and 5-16K in 1T measured by the standard four-probe technique are shown in Fig.3 and Fig.4. The transport critical current densities, $J_c$, are calculated from V-I curves by using a criterion of $1\mu v/cm$ and total superconducting cross section. The results are displayed in Fig.5. The transport $J_c$ up to $80,000A/cm^2$ (10K,0 T) and $13,600A/cm^2$ (10K,1T) are obtained. The results indicate that the intrinsic transport $J_c$ is high, and the practical $J_c$ will be improved with proper preparation process. Recently, we have fabricated the Cu-stabilized $MgB_2$ tape with Ta buffer layer and the initial measurement results (in CRTBT,CNRS France) show that the transport critical current is around 450A at 4.2K in null field. The farther measurement and analysis are underway.

**Conclusion**

The 18-filament $Cu/MgB_2$ tapes with NbZr buffer layer were prepared by the PIT technique. The tapes were sintered at 600-1000℃ for 1-10h in a high purity argon. XRD result revealed that the main phase is $MgB_2$. The transport critical current measurements were performed using a standard four-probe technique and high transport critical current density of $80,000A/cm^2$ (10K,0 T) and $13,600A/cm^2$ (10K,1T) have been obtained.

Fig.1: X-ray diffraction patterns of 18-filament Cu/NbZr/MgB$_2$ tape.

Fig.2: The optical photography of 18-filament Cu/NbZr/MgB$_2$ composite.

Fig.3: The V-I characteristics curves of multifilament tape at different temperatures and 0T.

Fig.4: The V-I characteristics curves of multifilament tape at different temperatures and 1T.

Fig.5: The magnetic field dependence of J$_c$ at different temperatures in 0T and 1T.

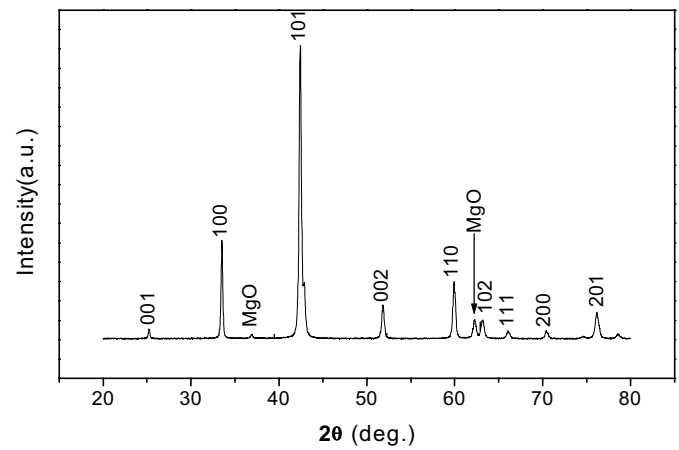

**Fig.1**

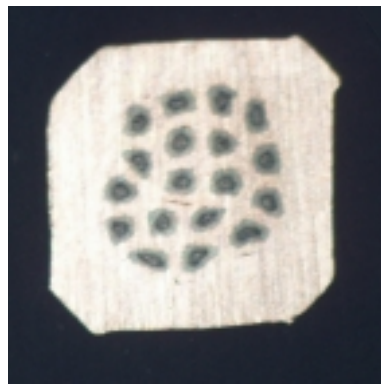

**Fig.2**

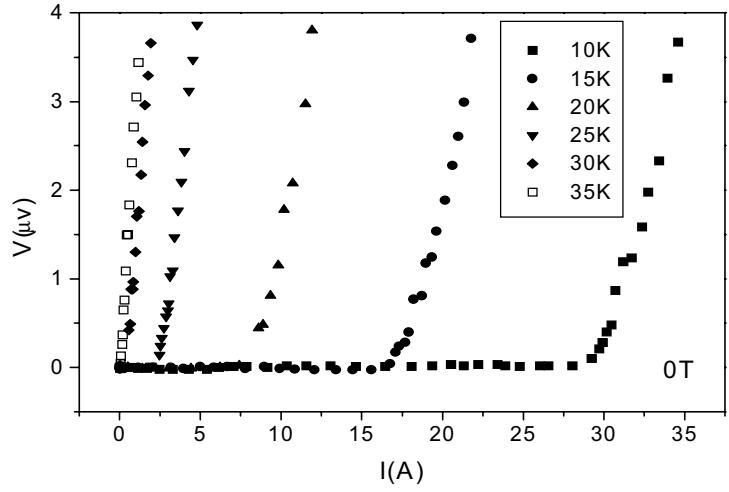

**Fig.3**

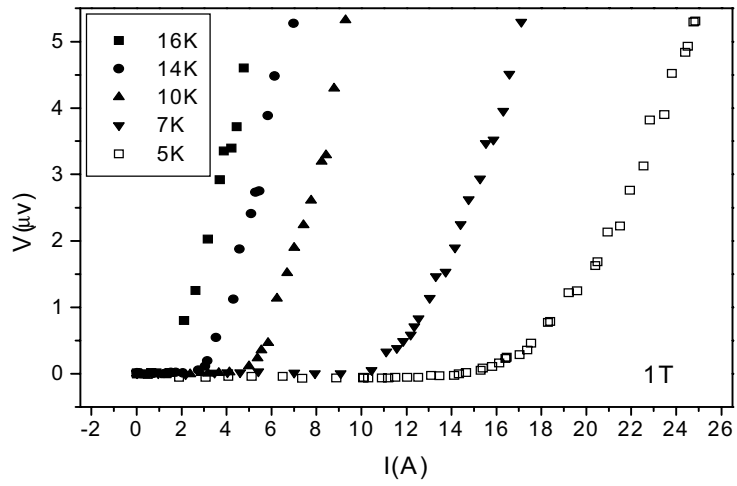

**Fig.4**

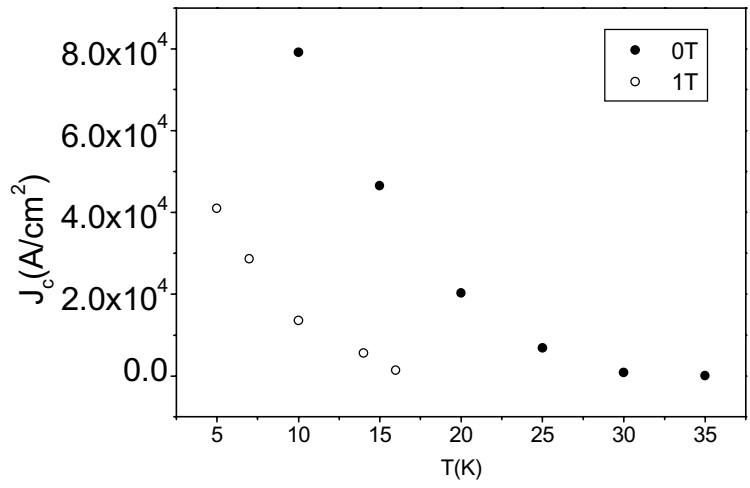

**Fig.5**